\documentclass[12pt]{article} 
\pdfoutput=1
\usepackage{amssymb,graphicx}
\usepackage{epstopdf}
\usepackage{amsmath,amsfonts}
\usepackage{epsfig} 
\usepackage{graphicx,graphics}
\usepackage{xcolor}
\usepackage{hyperref}

\begin{document} 

\sloppy
\title{\bf Reconciling spontaneous scalarization with cosmology}
%\title{\bf Taming the cosmological instability in scalar-tensor theories exhibiting scalarization }
\author{Timothy Anson$^{a}$, Eugeny Babichev$^{a,b}$, and Sabir Ramazanov$^{c}$\\
 \small{$^a$\em Laboratoire de Physique Th\'eorique, CNRS,} \\ 
 \small{\em  Univ. Paris-Sud, Universit\'e Paris-Saclay, 91405 Orsay, France}\\
 \small{$^b$\em Sorbonne Universit\'e, CNRS, UMR7095, Institut d'Astrophysique de Paris, ${\mathcal{G}}{\mathbb{R}}\varepsilon{\mathbb{C}}{\mathcal{O}}$,}\\
\small{\em 98bis boulevard Arago, F-75014 Paris, France}\\
\small{$^c$ \em CEICO, Institute of Physics of the Czech Academy of Sciences,}\\
\small{\em Na Slovance 1999/2, 182 21 Prague 8, Czech Republic}
 }

{\let\newpage\relax\maketitle}

\begin{abstract}
We discuss the Damour--Esposito-Far\`ese model of gravity, which predicts the spontaneous scalarization of neutron stars in a certain range of parameter space. In the cosmological setup, the scalar field responsible for scalarization is subject to a tachyonic instability during inflation and the matter domination stage, resulting in a large value of the field today.
This value feeds into 
the parametrized post-Newtonian parameters, which turn out to be in gross conflict with the Solar system measurements. 
We modify the original Damour--Esposito-Far\`ese model by coupling the scalar to the inflaton field.
This coupling acts as an effective mass for the scalar during inflation. 
For generic couplings that are not extremely small,
the scalar (including its perturbations) relaxes to zero with an exponential accuracy by the beginning of the hot stage. While the scalar exhibits growth during 
the subsequent cosmological stages, the resulting present value remains very small---in a comfortable agreement with the Solar system tests.
\end{abstract}

\section{Introduction}
Certain scalar-tensor theories of gravity result in the spontaneous scalarization of compact objects---black holes and/or neutron stars~\cite{Damour:1993hw, Damour:1996ke, Silva:2017uqg, Doneva:2017bvd, Antoniou:2017acq}. 
The essence of scalarization is in the amplification of the scalar field in the vicinity of compact objects relative to its 
cosmological value. Consequently, 
one expects predictions in the strong gravity regime to differ from the ones of General Relativity (GR), even if deviations from GR are unobservable in the weak gravity and quasi-static regimes, e.g., in the Solar system. That situation 
is exemplified by the Damour--Esposito-Far\`ese (DEF) model of scalarization~\cite{Damour:1993hw, Damour:1996ke}, which is the main focus of the present work. 

The key ingredient underlying models exhibiting scalarization is the tachyonic instability 
experienced by the scalar field due to its coupling to the curvature invariants. 
The equation of motion for the scalar field always has a trivial solution $\varphi=\mbox{const}$. 
However, the latter is subject to a tachyonic instability, which triggers the 
appearance of scalar hair---a non-trivial profile of the scalar field in the vicinity 
of compact objects. In the present work, we will study cosmological manifestations of the tachyonic instability. As a result of the latter, the scalar field responsible for scalarization 
has runaway solutions in certain cosmological backgrounds. For example, in the model with the scalar coupled to the Gauss--Bonnet curvature~\cite{Silva:2017uqg, Doneva:2017bvd,Antoniou:2017acq}, 
there is a catastrophic instability developed during the inflationary stage~\cite{Anson:2019uto}. 
Possible ways to resolve the problem in the model with the Gauss-Bonnet invariant fail essentially because of the 
huge (from the point of view of particle physics) dimensionful coupling constant needed to give rise to scalarization of astrophysical objects.
On the contrary, the original model of scalarization by Damour and Esposito-Far\`ese does not contain extra dimensionful parameters, and the only additional dimensionless constant is of order of unity. This feature makes the DEF model attractive from the perspective of physically viable modifications.

%Proposed ways of solving the problem encounter the  strong coupling issue.

In the DEF model of scalarization, the cosmological tachyonic instability occurs whenever the trace of the total matter energy density is larger than zero, 
i.e., always except during the radiation-dominated stage. Unless the scalar field is tuned to zero with high accuracy at the 
onset of the matter-dominated stage, it grows to large values by the present day---in conflict with the 
Solar system tests~\cite{Sampson:2014qqa, Alby:2017dzl, Anderson:2016aoi}. In this work, we propose a modification of the original DEF scenario where this tuning is automatic. This is achieved by coupling the field $\varphi$ to the inflaton $\chi$, i.e., $\sim \varphi^2 \chi^2$. 
This coupling gives an effective mass term for $\varphi$ during inflation. For generic super-Planckian values of the inflaton and coupling constants that are not extremely small, the mass is larger than the inflationary Hubble rate. As a result, the field $\varphi$ relaxes to an 
exponentially small value. Note that upon the inflaton decay, the effective mass for $\varphi$ vanishes. Consequently, 
at post-inflationary times the model of interest reduces to the original DEF scenario.

This paper is organized as follows. In Section~2, we review the DEF model of scalarization. In Section~3, we discuss the cosmological tachyonic instability 
which leads to the conflict with Solar system tests. We propose a modification of the DEF model, in which the conflict is resolved, 
in Section~4. We conclude in Section~5 with discussions.

\section{The Damour--Esposito-Far\`ese model of scalarization}

We use the same notations as in the original work on scalarization~\cite{Damour:1993hw}, but assume the $(+,-,-,-)$ signature for the metric. We begin with the Einstein frame action:
\begin{equation}
\label{DEFEinstein}
S_{\text{E}}=\int \text{d}^4 x \frac{\sqrt{-g}}{2\kappa} \left[-R+2\partial_{\mu} \varphi \partial^{\mu} \varphi- 2V(\varphi) \right]+S_\text{m} \left[A^2 (\varphi) g_{\mu \nu}, \psi_\text{m} \right] \; ,
\end{equation}
where $\kappa =8\pi G$, $G$ is Newton's constant, $\psi_\text{m}$ is the collective notation for matter fields, and the function $A(\varphi)$ is defined as 
\begin{equation}
\label{coupling}
A(\varphi)=e^{\frac{1}{2} \beta \varphi^2} \; ,
\end{equation}
where $\beta$ is a constant, which feeds into deviations from GR. Following the notations of Ref.~\cite{Damour:1993hw}, we have chosen the field $\varphi$ to be dimensionless.
%The presence of the quadratic term in the exponent, $\beta\varphi^2$, with negative $\beta$, results in scalarization of neutron stars if $\beta\lesssim -4$.
 In the original DEF model, the potential 
$V (\varphi)$ is absent. We keep it, however, because it plays a crucial role in our discussion later on. 
It is worth mentioning that the action~(\ref{DEFEinstein}) implies the universal coupling of all the matter fields to the metric. 
We proceed with this assumption in the bulk of the paper. There are, however, alternative options, e.g., of an inflaton coupled to the Einstein metric $g_{\mu\nu}$ differently than other matter fields. 
Though such a coupling appears to be rather unnatural, we comment on this possibility in the concluding Section~\ref{sec:Discussion}.

In the Jordan frame $\tilde{g}_{\mu \nu}=A^2 (\varphi) g_{\mu \nu}$, where the 
matter fields follow geodesics, the equivalent action is given by 
\begin{equation}
\nonumber 
S_{\text{J}}=\int \text{d}^4 x \frac{\sqrt{-\tilde{g}}}{2\kappa} \left[-\tilde{\varphi} \tilde{R}+\frac{\omega (\tilde{\varphi})}{\tilde{\varphi}} \tilde{g}^{\mu \nu} \partial_{\mu} \tilde{\varphi} \partial_{\nu} \tilde{\varphi} -\Pi (\tilde{\varphi})\right]+S_\text{m} \left[\tilde{g}_{\mu \nu}, \psi_\text{m} \right] \; ,
\end{equation}
where 
\begin{equation}
\label{rel}
\left(\frac{\text{d} \ln A}{\text{d} \varphi} \right)^2 =[2\omega (\tilde{\varphi})+3]^{-1} \qquad A^2 (\varphi)=\frac{1}{\tilde{\varphi}}\; .
\end{equation}
The potential $\Pi (\tilde{\varphi})$ is related to $V(\varphi)$ by 
\begin{equation}
\nonumber 
\Pi (\tilde{\varphi}) =2 \tilde{\varphi}^2 V(\varphi (\tilde{\varphi})) \; .
\end{equation}
This potential is zero in the DEF model. 

The essence of scalarization is as follows. The equation of motion for the field $\varphi$ derived from the action~(\ref{DEFEinstein}) reads,
\begin{equation}
\label{tachyonic} 
\square \varphi+\frac{\kappa}{2} \alpha (\varphi) T^{\text{m}}+\frac{1}{2} \frac{\partial V}{\partial \varphi}=0 \; ,
\end{equation}
where $\alpha(\varphi) \equiv \frac{\text{d} \ln A(\varphi)}{\text{d} \varphi} =\beta \varphi$ plays the role of the coupling constant to the matter fields; the matter stress-energy tensor is defined as 
$T^{\text{m}}_{\mu \nu}=\frac{2}{\sqrt{-g}} \frac{\delta S_{\text{m}}}{\delta g^{\mu \nu}}$ and its trace as $T^{\text{m}}=g^{\mu \nu} T^{\text{m}}_{\mu \nu}$. One can see that $\varphi=0$ solves this equation for the potential $V(\varphi)=0$. 
For $\beta <0$ and $T^{\text{m}}>0$, the scalar acquires a tachyonic effective mass, which hints at the existence of other, stable solutions 
of Eq.~\eqref{tachyonic}. This is indeed the case for $\beta \lesssim -4$ in the strong gravity regime inside neutron stars~\cite{Damour:1993hw}. Namely, the field $\varphi$ acquires a non-trivial profile
which matches the constant cosmological value $\varphi_0 \equiv \varphi (t_0)$, where $t_0 \approx 13.8~\cdot 10^{9}~\mbox{years}$ is the present time. 

In the model~\eqref{DEFEinstein}, parametrized post-Newtonian (PPN) parameters are given by~\cite{Damour:1992we}
\begin{equation}
\label{PPN}
\gamma_{\text{PPN}}-1=\frac{-2 \alpha^2 (\varphi_0)}{1+\alpha^2 (\varphi_0)} \qquad \beta_{\text{PPN}}-1 =\frac{\beta \alpha^2 (\varphi_0)}{\left[1+\alpha^2 (\varphi_0)\right]^2} \; .
\end{equation}
In the limit $\alpha (\varphi_0) \rightarrow 0$, the PPN parameters coincide with those of GR. This limit corresponds to $\varphi_0 \rightarrow 0$. 
Using the constraint on the PPN parameter $\gamma_{\text{PPN}}$ from the Shapiro time-delay measurement: $\gamma_{\text{PPN}}=1 \pm (2.1 \pm 2.3) \times 10^{-5}$ given in Ref.~\cite{Bertotti:2003rm}, we get for $|\beta| \simeq 4$ the following upper bound on $\varphi_0$:
\begin{equation}
\label{closetozero}
\varphi_0 \lesssim 10^{-3} \; .
\end{equation}
For these values, the DEF model is indistinguishable from GR in the weak field 
and quasi-static regimes. However, even with a vanishing value of the field $\varphi$ at cosmological scales, neutron stars experience scalarization, 
leading to testable deviations from GR in the strong field regime~\cite{Freire:2012mg, Shao:2017gwu}.
On the other hand, as we discuss in the next section, in the original DEF model with $V(\varphi)=0$, the values~\eqref{closetozero} are non-realistic. Indeed, the tachyonic instability 
triggers runaway cosmological solutions for the field $\varphi$, so that  $\varphi_0 \gg 1$---in direct conflict with the Solar system constraints~\cite{Sampson:2014qqa, Alby:2017dzl, Anderson:2016aoi}.

\section{Setting the problem: cosmological instability of the field $\varphi$}
In the present section, we estimate the effect of the tachyonic instability in the DEF scenario. 
The presence of the instability is  evident from Eq.~\eqref{tachyonic}, and it has the same origin as the instability responsible 
for the  scalarization of neutron stars. If $V(\varphi) =0$ as in the original DEF scenario, the second term on the l.h.s. of Eq.~\eqref{tachyonic} mimics the mass term. Apart from the radiation-dominated stage, when 
$T^{\text{m}}=0$ approximately, this mass term is negative for $\beta <0$ and thus leads to the tachyonic instability. Let us estimate the rate of this instability during the matter-dominated stage. Neglecting backreaction 
of the scalar $\varphi$ on the metric, from Eq.~\eqref{tachyonic} one obtains, 
\begin{equation}
\nonumber 
\ddot{\varphi}+3H \dot{\varphi}+\frac{3}{2} \beta H^2 \varphi =0 \; .
\end{equation}
Recall that we work in the Einstein frame. Hence, the scale factor $a(t)$ and the Hubble expansion rate $H(t)$ are defined in this frame.
However, 
in what follows we will not make a distinction between the energy-momentum tensor in the two frames, 
since $T_{\mu\nu}^\text{m}\simeq\tilde{T}_{\mu\nu}^\text{m}$ as long as $\varphi \ll 1$. Later on, we will see that $\varphi$ is indeed extremely close to zero in our scenario, so this assumption is justified.
The above equation has the growing solution given by
\begin{equation}
\nonumber 
\varphi \simeq \varphi_{\text{eq}} \left(\frac{t}{t_{\text{eq}}} \right)^{\frac{\sqrt{1-\frac{8\beta}{3}}-1}{2}} \; ,
\end{equation}
where $H=\frac{2}{3t}$ and the subscript 'eq' denotes the matter-radiation equality. From this relation, one can convert the upper bound on $\varphi_0$ in Eq.~\eqref{closetozero} into a limit on $\varphi_{\text{eq}}$. We substitute $t_{\text{eq}} \approx 5\cdot10^4~\mbox{years}$, $t_0 \approx 13.8 \cdot 10^{9}~\mbox{years}$, $\beta=-4.5$, and obtain 
\begin{equation}
\label{upper}
\varphi_{\text{eq}} \lesssim 10^{-10} \; .
\end{equation} 
Note that we assumed the matter-dominated stage continues up to the present day, but taking into account the current accelerated expansion of the Universe does not alter this estimate considerably. 

Strictly speaking, the tachyonic instability is also present during the radiation-dominated stage. The reason is that $T^{\text{m}}$ is slightly different from zero mainly 
due to the Dark Matter contribution. However, this instability is very mild, as we will show explicitly. The equation governing evolution of the field $\varphi$ during 
the radiation-dominated stage is given by 
\begin{equation}
\nonumber 
\ddot{\varphi}+3H \dot{\varphi}+\frac{\kappa}{2} \beta \varphi \rho_{\text{matter}} =0 \; .
\end{equation}
The non-relativistic matter energy density evolves as 
\begin{equation}
\nonumber 
\rho_{\text{matter}} (t)=\rho_{\text{matter}, \text{eq}}  \cdot \frac{a^3_{\text{eq}}}{a^3(t)}  \; .
\end{equation}
We estimate $\rho_{\text{matter}, \text{eq}}$ as 
\begin{equation}
\nonumber 
\rho_{\text{matter}, \text{eq}} =\rho_{\text{rad}, \text{eq}} \simeq \frac{3H^2 (t_{\text{eq}})}{\kappa} \; ,
\end{equation}
where $H (t_{\text{eq}}) \simeq \frac{1}{2t_{\text{eq}}}$ is the Hubble rate at equality obtained by extrapolating the Hubble rate $H(t)=\frac{1}{2t}$ during radiation-domination; $\rho_{\text{rad}, \text{eq}}$ 
is the radiation energy density at equality. Putting everything together and substituting the scale factor $a(t) \propto \sqrt{t}$, we obtain the equation
\begin{equation}
\nonumber 
\ddot{\varphi}+\frac{3}{2t} \dot{\varphi}+\frac{3\beta}{8t^2_{\text{eq}}} \left(\frac{t_{\text{eq}}}{t} \right)^{3/2}\varphi=0 \; .
\end{equation}
We have checked that it has the growing solution:
\begin{equation}
\nonumber 
\varphi \simeq 2\varphi_\text{i} \frac{ I_{1} (\sqrt{6|\beta|} \xi^{1/4})}{\sqrt{6|\beta|} \xi^{1/4}} \; ,
\end{equation}
where $I_1$ is the modified Bessel function of the first kind of order 1, $\xi \equiv \frac{t}{t_{\text{eq}}}$, and $\varphi_\text{i}$ is the value of the field at the onset of the radiation-dominated stage, i.e., in the formal limit $t \rightarrow 0$. Substituting known values of $I_1$, one obtains
\begin{equation}
\nonumber 
\frac{\varphi_{\text{eq}}}{\varphi_\text{i}} \simeq 10 \; .
\end{equation}
Combining with Eq.~\eqref{upper}, 
we conclude that $\varphi_\text{i}$ is constrained as 
\begin{equation}
\label{constr}
\varphi_\text{i} \lesssim 10^{-11} \; .
\end{equation} 
This means that to achieve consistency with Solar system tests, the post-inflationary value of $\varphi$ should be tuned to zero with high accuracy. 
Note that the value $\varphi_\text{i}$ 
is also subject to BBN constraints. However, the latter are very weak~\cite{Alby:2017dzl}, typically $\varphi_\text{i}\lesssim 1$. Hence, once we manage to 
satisfy the constraint~\eqref{constr}, the BBN limit will be automatically obeyed.

One comment is in order here. We have assumed that the field $\varphi$ is homogeneous. In practice, there are small inhomogeneities due to cosmological perturbations imposed on the 
field $\varphi$. These inhomogeneities evolve differently depending on their characteristic wavelength. Namely, there is an upper bound on the wavenumber of cosmological 
modes which experience the instability:
\begin{equation}
\label{constraint}
\frac{k}{a (t_{\text{eq}})} \lesssim H (t_{\text{eq}}) \; .
\end{equation}
Indeed, spatial inhomogeneities of the field $\varphi$ characterized by the wavenumber $k$ yield the term $\sim \frac{k^2}{a^2} \varphi_{{\bf k}}$ in the evolution 
equation of the corresponding mode $\varphi_{{\bf k}}$:
\begin{equation}
\label{mdk}
\ddot{\varphi}_{{\bf k}}+3H \dot{\varphi}_{{\bf k}}+\frac{3}{2}\beta H^2 \varphi_{{\bf k}}+...=0 \; .
\end{equation}
Here the ellipses stand for the terms sourced by the gravitational potential and matter energy density perturbations, which give a negligible contribution. For perturbations violating the upper bound~\eqref{constraint}, the second term in Eq.~\eqref{mdk} screens the term ${\cal O} (H^2)$, which would otherwise give rise to the tachyonic instability. As a result, short wavelength modes decay as $\varphi_{{\bf k}} \propto \frac{1}{a}$, as it should be 
for the case of a massless scalar field in the expanding Universe. Thus, 
we will focus on perturbations obeying Eq.~\eqref{constraint} in what follows. 

The present work aims to explain the small value $\varphi_\text{i}$ constrained by Eq.~\eqref{constr}. This problem is exacerbated during the inflationary stage, when the field $\varphi$ also 
experiences the tachyonic instability. We show in the Appendix that even if classically the field $\varphi$ is set to zero at the onset of inflation, its vacuum fluctuations get largely 
amplified beyond the horizon, quickly shifting $\varphi$ from zero to $\varphi \gg 1$. The latter is not only inconsistent with the Solar system tests but also with the existence of the inflationary 
stage\footnote{A similar issue is present for models with scalarization due to the scalar-Gauss-Bonnet coupling. However, the instability there is even stronger~\cite{Anson:2019uto}.}. Note that according to Eq.~\eqref{rel}, $\varphi \gg 1$ 
corresponds to a huge $\tilde{\varphi} \ggg 1$ in the Jordan frame. Hence, the field $\tilde{\varphi}$ quickly comes to dominate the evolution of the Universe, and inflation terminates. We conclude that the DEF scenario should be modified at least in the very early Universe, and one modification of this type is discussed in the next section.

Before that, let us briefly comment on the solutions of the problem of the tachyonic instability existing in the literature. In Ref.~\cite{Alby:2017dzl}, 
it was proposed to endow the scalar with a small mass $m$ by promoting the potential $V(\varphi)$ to $V(\varphi) =\frac{m^2 \varphi^2}{2}$. As the Hubble rate drops down to $H \simeq m$, the field $\varphi$ starts to decay oscillating 
about the minimum of its potential at $\varphi = 0$. From this point on, it contributes to the Dark Matter content of the Universe. Given post-inflationary conditions for the field $\varphi$ assumed in Ref.~\cite{Alby:2017dzl}, 
i.e., $\varphi_\text{i} \simeq 1$ and $\dot{\varphi}_\text{i} \simeq 0$, 
the mass $m$ should be extremely tiny, i.e., $m \lesssim 10^{-28}~\mbox{eV}$. For masses violating this bound, the field $\varphi$ gives an unacceptably large contribution 
to the energy density of the Universe. Apart from tuning the mass $m$, the
instability during inflation remains an issue, as discussed above and in the Appendix. As a result of this instability, one should expect the initial condition $\varphi_\text{i} \gg 1$ rather than $\varphi_\text{i} \simeq 1$.

In passing, we would like to point out that the instability during inflation and at later stages can be avoided by promoting the function $\ln A(\varphi)$ to~\cite{Anderson:2016aoi}
\begin{equation}
\label{quartic}
\ln A(\varphi)=\frac{\beta \varphi^2}{2}+\frac{\lambda \varphi^4}{4} \; .
\end{equation}
Choosing the extra parameter $\lambda>0$, one can stabilize the field $\varphi$ during inflation, so that it evolves 
close to the effective minimum $\varphi=\sqrt{-\beta/\lambda}$ right until present. Unfortunately, this scenario does not work, 
because with $\varphi_0 \neq 0$ and $\lambda \neq 0$, the scalarization of neutron stars does not occur. 

In this work, we follow another approach to the problem of consistency with Solar system tests. Namely, we will find a way to relax the field $\varphi$ to 
tiny values during inflation, well below the upper bound in Eq.~\eqref{constr}, while at the same time retaining the original form of the DEF model at post-inflationary times.  

\section{Cosmological relaxation of the field $\varphi$ to zero}

The idea is to couple the field $\varphi$ to the inflaton $\chi$, i.e., consider the interaction of the form $\sim \varphi^2 \chi^2$. Such a coupling induces a large effective mass for the field $\varphi$ during inflation, so that $\varphi$ relaxes to an exponentially small value. 
The effective mass term vanishes upon the inflaton decay, so that we end up with the standard DEF scenario after inflation. While the tachyonic instability during the matter-dominated stage is still present, there is not enough time for the field $\varphi$ 
to grow to large values by cosmological mechanisms. 
In other words, the inequality $\varphi_0 \ll 10^{-3}$ is always satisfied---in agreement with the Solar system tests. 

We assume that inflation is driven by the canonical scalar field $\chi$ rolling down the slope of its (almost) flat potential $U(\chi)$. In the Einstein frame its action is given by
\begin{equation}
\nonumber 
S_\text{m} \left[A^2 (\varphi)g_{\mu \nu}, \psi_\text{m} \right] \rightarrow S_{\chi} \left[A^2 (\varphi)g_{\mu \nu}, \chi \right]=\int \text{d}^4 x \sqrt{-\tilde{g}} \left[\frac{1}{2} \tilde{g}^{\mu \nu} \partial_{\mu} \chi \partial_{\nu} \chi-U(\chi) \right] \left. \right|_{\tilde{g}_{\mu \nu}=A^2(\varphi) g_{\mu \nu}} \; .
\end{equation}
Note that unlike the field $\varphi$, the inflaton $\chi$ is assumed to have a canonical mass dimension.
We modify the DEF model by assuming a non-zero interacting potential $V$:
\begin{equation}
\label{inter}
V(\varphi) \rightarrow V(\varphi, \chi)=g^2 \varphi^2 \chi^2 \; ,
\end{equation} 
where $g^2$ is some dimensionless coupling. Thus the field $\varphi$ has the effective mass $g^2 \chi^2$ due to the coupling to the inflaton. 
We require that 
\begin{equation}
\label{hierarchy}
g^2 \chi^2 \gg H^2 \; .
\end{equation}
Namely, the field $\varphi$ is effectively superheavy meaning that its effective mass is larger than the inflationary Hubble rate (but still below the Planckian scale). In this case, $\varphi$ relaxes to zero within a few Hubble times. For typical values 
$\chi \simeq M_{Pl}$ and $H \simeq 10^{13}~\mbox{GeV}$, the constant $g^2$ can be as small as $g^2 \simeq 10^{-12}$. 
Hence, the mechanism which cures the instabilities can operate in a very weakly coupled regime. In the Jordan frame, the potential~\eqref{inter} is transformed to 
\begin{equation}
\nonumber 
\Pi (\tilde{\varphi}, \chi)=2g^2 \varphi^2 (\tilde{\varphi}) \tilde{\varphi}^2\chi^2 \qquad \varphi^2 (\tilde{\varphi})=-\frac{\ln \tilde{\varphi}}{\beta} \; .
\end{equation}
Note that Eq.~\eqref{rel} implies $\tilde{\varphi}>1$ for $\beta<0$. Hence, the Jordan frame interacting potential $\Pi (\tilde{\varphi}, \chi)$ is positive. We see that modulo the logarithmic correction, the interacting potential has a quadratic form in the Jordan frame as well. 
Therefore it is not important in which frame the coupling to the inflaton is introduced.
We now list the set of equations relevant for future purposes. Einstein-Hilbert equations are given by 
\begin{equation}
\nonumber 
R_{\mu \nu}-\frac{1}{2}g_{\mu \nu} R=\kappa T^{\chi}_{\mu \nu}+T^{\varphi}_{\mu \nu} \; ,
\end{equation}
where 
\begin{equation}
\nonumber 
T^{\varphi}_{\mu \nu}=2\partial_{\mu} \varphi \partial_{\nu} \varphi -g_{\mu \nu}\partial_{\alpha} \varphi \partial^{\alpha} \varphi +g_{\mu \nu} V(\varphi, \chi) \; ,
\end{equation}
and 
\begin{equation}
T^{\chi}_{\mu \nu}=A^2(\varphi) \partial_{\mu} \chi \partial_{\nu} \chi -\frac{1}{2}g_{\mu \nu} A^2 (\varphi)\partial_{\alpha} \chi \partial^{\alpha} \chi +g_{\mu \nu} A^4(\varphi)U(\chi) \; .
\end{equation}
Note that the indices are raised and lowered with the Einstein metric $g_{\mu \nu}$
The equations of motion for the field $\varphi$ and the inflaton are given by Eq.~\eqref{tachyonic}, where $T^{\text{m}}$ is replaced by $T^{\chi}$, and  
\begin{equation}
\nonumber 
\tilde{\square} \chi+U_{\chi}+\frac{1}{\kappa A^4(\varphi)} V_{\chi} (\varphi, \chi)=0 \; ,
\end{equation}
respectively. 
%which can be rewritten as 
%\begin{equation}
%\nonumber 
%g^{\mu \nu} \partial_{\mu} \partial_{\nu} \chi+g^{\mu \nu}\partial_{\mu} \ln \sqrt{-g} \partial_{\nu} \chi+2\alpha (\varphi) g^{\mu \nu} \partial_{\mu} \chi \partial_{\nu} \varphi+\partial_{\mu} g^{\mu \nu} \partial_{\nu} \chi+A^2(\varphi) U_{\chi} +%\frac{1}{\kappa A^2(\varphi)} V_{\chi} (\varphi, \chi)=0 \; . 
%\end{equation}

\subsection{Relaxing the background value of $\varphi$ to zero} 
The Friedmann equation is given by 
\begin{equation}
\nonumber 
3H^2 =\dot{\varphi}^2+V(\varphi, \chi)+\frac{\kappa}{2} A^2 (\varphi) \dot{\chi}^2+\kappa A^4(\varphi) U(\chi) \; .
\end{equation}
%The derivative of the Hubble rate is given by
%\begin{equation}
%\nonumber 
%-2\dot{H}=\kappa A^2(\varphi)\dot{\chi}^2+2\dot{\varphi^2} \; .
%\end{equation}
The background evolution of the scalar $\varphi$ is governed by the equation
\begin{equation}
\label{backvarphi}
\ddot{\varphi}+3H \dot{\varphi}+\frac{\kappa}{2} \alpha (\varphi) \cdot \left[4A^4(\varphi) U(\chi)-A^2 (\varphi) \dot{\chi}^2 \right]+g^2 \chi^2 \varphi =0 \; .
\end{equation}
As usual, we assume that the inflaton potential dominates the energy density of the Universe, i.e., $3H^2 \approx \kappa A^4(\varphi) U(\chi)$. Consequently, we drop the second term in the square brackets of 
Eq.~\eqref{backvarphi}. The background equation for $\varphi$ simplifies to 
\begin{equation}
\nonumber 
\ddot{\varphi}+3H \dot{\varphi}+m^2 \varphi =0 \; ,
\end{equation}
where $m^2$ is the full effective mass of the field $\varphi$ defined by
\begin{equation}
\nonumber
m^2=g^2 \chi^2 +6\beta H^2 \; .
\end{equation}
Provided that the condition~\eqref{hierarchy} is obeyed and $|\beta|$ is not very large, the field $\varphi$ evolves as a superheavy field, which relaxes to zero within a few Hubble times. In the exact de Sitter space-time approximation, 
the solution for the field $\varphi$ is given by
\begin{equation}
\nonumber 
\varphi =\frac{C}{a^{3/2}} \cdot \cos \left[\sqrt{m^2-\frac{9H^2}{4}}\,t +\delta \right] \; ,
\end{equation}
where $C$ and $\delta$ are irrelevant constants. We conclude that starting from subplanckian values $\varphi < 1$, by the end of inflation the field $\varphi$ is relaxed to 
\begin{equation}
\nonumber 
\varphi \lesssim 10^{-39} \; ,
\end{equation}
where the upper bound corresponds to the minimal duration of inflation---about $60$ e-foldings. Generically, the duration of inflation is much larger, so one can safely set the background value of $\varphi$ to zero. 

The background evolution of the inflaton is governed by the equation: 
\begin{equation}
\nonumber 
\ddot{\chi}+3H \dot{\chi} +2\alpha (\varphi) \dot{\chi} \dot{\varphi} +A^2 (\varphi) U_{\chi}+\frac{1}{\kappa A^2(\varphi)} V_{\chi} (\varphi, \chi)=0 \; .
\end{equation}
As $\varphi \rightarrow 0$, one has $\alpha (\varphi) \rightarrow 0$, $A(\varphi) \rightarrow 1$, and $V_{\chi} \rightarrow 0$. Therefore, the evolution of the inflaton proceeds as in GR. 

\subsection{Relaxing the perturbations $\delta \varphi$ to zero} 
One may naively expect the field $\varphi$ to develop superhorizon perturbations $\delta \varphi \simeq \frac{H}{M_{Pl}}$ 
for each mode. Taking into account that for standard inflation scenarios $\frac{H}{M_{Pl}}\sim 10^{-6}$, such perturbations would be a problem for the DEF scenario, cf. Eq.~(\ref{constr}).
Such a situation would occur for light fields during inflation. However, our case is different, as the field $\varphi$ is effectively superheavy. 
Below we prove rigorously that perturbations $\delta \varphi$, which source the present day cosmological value of $\varphi$, are 
exponentially suppressed by the end of inflation. 

In the Newtonian gauge linear metric perturbations are given by
\begin{equation}
\nonumber 
\text{d}s^2 =(1+2\Phi)\text{d}t^2 -a^2 (1-2\Psi) \delta_{ij} \text{d}x^{i} \text{d}x^{j} \; .
\end{equation}
In the absence of the anisotropic stress, which is the case here, $\Phi=\Psi$. 
%Then, the $0i$-component of linearized Einstein's equations takes the form 
%\begin{equation}
%\nonumber 
%\dot{\chi}+H \chi=4\pi G A^2 (\varphi) \dot{\chi} \delta \chi +\dot{\varphi} \delta \varphi \; .
%\end{equation}
We are primarily interested in the linear perturbation $\delta \varphi$. The relevant equation is given by
\begin{equation}
\label{eqvarphipert}
\begin{split}
&\delta \ddot{\varphi}-\frac{1}{a^2} \partial_i \partial_i \delta \varphi-2 \ddot{\varphi} \Phi-4 \dot{\varphi} \dot{\Phi} -6 H \dot{\varphi} \Phi +3H \delta \dot{\varphi} +\frac{\kappa}{2} \alpha (\varphi) \delta T^{\text{m}} +\\ 
&+\frac{\kappa}{2} \frac{\partial \alpha (\varphi)}{\partial \varphi} T^{\text{m}} 
\delta \varphi+\frac{1}{2} \frac{\partial^2 V}{\partial \varphi^2} \delta \varphi+\frac{1}{2} \frac{\partial^2 V}{\partial \varphi \partial \chi} \delta \chi=0 \; ,
\end{split}
\end{equation}
where 
\begin{equation}
\nonumber \delta T^{\text{m}}=16 A^4(\varphi) \alpha (\varphi) U(\chi) \delta \varphi+4 A^4 (\varphi) \frac{\text{d} U}{\text{d} \chi} \delta \chi -2A^2 (\varphi) \alpha (\varphi) \dot{\chi}^2 \delta \varphi +2A^2 (\varphi) \dot{\chi}^2 \Phi-2A^2(\varphi) \dot{\chi} 
\delta \dot{\chi} \; .
\end{equation}
While this equation looks rather complicated, it is simplified upon substituting the background value $\varphi=0$. We obtain in terms of the Fourier modes $\delta \varphi_{{\bf k}}$:
\begin{equation}
\nonumber 
\delta \ddot{\varphi}_{{\bf k}} +3H \delta \dot{\varphi}_{{\bf k}} +\frac{k^2}{a^2} \delta \varphi_{{\bf k}}+\frac{\kappa}{2} \frac{\partial \alpha (\varphi)}{\partial \varphi} T^{\text{m}} 
\delta \varphi_{{\bf k}}+\frac{1}{2} \frac{\partial^2 V}{\partial \varphi^2} \delta \varphi_{{\bf k}}=0 \; .
\end{equation}
This is a homogeneous equation, which describes a damped oscillator with an almost constant large mass. The modes $\delta \varphi_{{\bf k}}$ decay as $\frac{1}{a^{3/2}}$ in the superhorizon regime. 
Hence, they have negligibly small amplitudes by the end of the inflationary stage. We will make an exact estimate of the field $\varphi$ due to its perturbations shortly.

Before going into details let us make two comments. First, note that the vanishing background value of $\varphi$ shields perturbations $\delta \varphi$ from the metric 
and inflaton fluctuations $\delta \chi$. Generally, the latter source adiabatic perturbations, which turn out to be zero in our case. This is also evident from the expression for adiabatic perturbations in the superhorizon regime~\cite{Polarski:1994rz}: 
\begin{equation}
\nonumber 
\frac{\delta \varphi_{\text{ad}}}{\dot{\varphi}}=\frac{\delta \chi}{\dot{\chi}}=\frac{1}{a} \cdot \left(C_1 \int^{t}_0 a \text{d}t'-C_2 \right) \; ,
\end{equation}
\begin{equation}
\nonumber 
\chi=C_1 \cdot \left( 1-\frac{H}{a} \int^{t}_0 a \text{d}t' \right) +C_2 \frac{H}{a} \; .
\end{equation}
Here $C_1$ and $C_2$ are some constants defined by the subhorizon evolution of the gravitational potential. Independently of their values, 
we have $\delta \varphi_{\text{ad}} \rightarrow 0$, because $\dot{\varphi} \rightarrow 0$. 

Second, we have considered only linear perturbations $\delta \varphi$. However, using the same argument as above one can show that 
once $\varphi \rightarrow 0$ and the linear perturbation $\delta \varphi \rightarrow 0$, the second order perturbation $\delta \varphi^{(2)}$ also obeys the 
homogeneous oscillator equation with the Hubble friction and a very large mass. Hence, it should also decay as $\delta \varphi^{(2)} \propto \frac{1}{a^{3/2}}$ in the superhorizon regime.

The above consideration shows that perturbations $\delta \varphi$ are indeed very small at the end of inflation. However, we still need to 
estimate the amplitude of perturbations in order to compare it with the constraint~(\ref{constr}).
We approximate inflation by an exact de Sitter stage and switch to the canonical variable $\delta \hat{\varphi}$ related to the original field $\delta \varphi$ by 
\begin{equation}
\label{canonical}
\delta \varphi=\sqrt{\frac{\kappa}{2}} \delta \hat{\varphi} \; .
\end{equation}
The solution for the field $\delta \hat{\varphi}$ obeying Bunch--Davies vacuum initial conditions is given by
\begin{equation}
\label{quantumfield}
\delta \hat{\varphi}=\int \frac{\text{d}^3 {\bf k}}{(2\pi)^{3/2}}\frac{\sqrt{\pi}}{2} H |\eta|^{3/2} \left[e^{\frac{\pi s}{2}} H^{(2)}_{is} (k|\eta|)  e^{-i {\bf kx}} A^{\dagger}_{{\bf k}}+e^{-\frac{\pi s}{2}} 
H^{(1)}_{i s} (k|\eta|) e^{i{\bf kx}} A_{{\bf k}} \right] \; ,
\end{equation}
where $\eta$ is the conformal time, $(A^{\dagger}_{{\bf k}}, A_{{\bf k}})$ is the pair of creation-annihilation operators, $H^{(1,2)}_{i s}$ are the Hankel functions of purely imaginary order~\cite{Bessel}; 
\begin{equation}
\nonumber 
s=\sqrt{\frac{m^2}{H^2}-\frac{9}{4}}  \; .
\end{equation}
Note that the Hankel functions $H^{(1,2)}_{i s}$ are not complex conjugate. 
Instead, the following relation is correct: $\left[H^{(1)}_{is} (k|\eta|) \right]^{*}=e^{\pi s} H^{(2)}_{is} (k|\eta|)$, which explains 
the presence of unconventional factors $e^{\frac{\pi s}{2}}$ in Eq.~\eqref{quantumfield}. For $s \gg 1$, one obtains in the limit $k |\eta | \rightarrow 0$, cf. Ref.~\cite{Bessel}:
\begin{equation}
\nonumber 
H^{(1,2)}_{i s} (k|\eta|)=\sqrt{\frac{2}{\pi s}} e^{\pm i s \ln \left[-\frac{1}{2} k\eta \right] \mp i\gamma_{s} \pm \frac{\pi s}{2}} \; ,
\end{equation}
where $\mp \gamma_{s}$ are irrelevant phases. The choice of the upper and the lower sign on the r.h.s. corresponds to the Hankel function of the first and the second kind, respectively. We are interested in the quantity $\langle \delta \tilde{\varphi}^2 \rangle_{\text{unstable}}$. The subscript 'unstable' means that 
we focus on the modes which are subject to the tachyonic instability during the matter-dominated stage. These modes have the cutoff $k_{\text{max}}$ defined by the condition~\eqref{constraint}. By the end of inflation, at the moment $\eta_f$, the dispersion 
$\langle \delta \tilde{\varphi}^2 \rangle_{\text{unstable}}$ is given by
\begin{equation}
\nonumber 
\langle \delta \hat{\varphi}^2 \rangle_{\text{unstable}} (\eta_f)=\frac{H^2}{12 \pi^2 s} \cdot \left|\frac{\eta_f}{\eta_{\times}} \right|^3 \; ,
\end{equation}
where $\eta_{\times}$ is defined by $k_{\text{max}} |\eta_{\times}| \simeq 1$. 

In terms of the original field $\varphi$, one finally gets 
\begin{equation}
\nonumber 
\langle \delta \varphi^2 \rangle_{\text{unstable}} (\eta_f) =\frac{H^2}{3\pi s M^2_{Pl}} \cdot \left|\frac{\eta_f}{\eta_{\times}} \right|^3 \; .
\end{equation}
Note that $\eta_{\times}$ roughly corresponds to $50$-$70$ e-foldings before the end of inflation, when cosmological modes exit the horizon. 
For the sake of concreteness, we assume $60$ e-foldings. Taking also $H \simeq 10^{-6} M_{Pl}$ (high scale inflation) and $s =10$, we find $\sqrt{\langle \delta \varphi^2 \rangle_{\text{unstable}} (\eta_f)} \simeq 10^{-46}$. 
The field $\varphi$ will be roughly frozen at this value during the radiation-dominated stage (modulo the factor '10' enhancement discussed in Section~3). At the matter-dominated stage and later it experiences the tachyonic instability. However, the resulting 
field $\varphi_0$ is still well below the upper bound, i.e., $\varphi_0 \lll 10^{-3}$---in a comfortable agreement with the Solar system tests.

\section{Discussions}
\label{sec:Discussion}

In the present work, we have proposed a way to extend the original DEF model of scalarization to cosmological scales, while retaining consistency with Solar system tests. In the cosmological context, the original model leads to a runaway solution for the relevant field $\varphi$, making the scenario 
inconsistent with existing PPN constraints unless the initial value of $\varphi$ is tuned to zero with high precision. We have found a modification 
of the original scenario in which this tuning is automatic. Namely, we have shown that if the field $\varphi$ responsible for scalarization is equipped with a coupling to the inflaton, it relaxes to zero with an exponential accuracy. Upon the inflaton decay, the coupling effectively vanishes, meaning that in our modified scenario all the predictions related to neutron stars are the same 
as in the original DEF model. 

Recall that in this work we have assumed the universal coupling of matter fields to the metric. 
Let us comment here on modifications of the model where the coupling is non-universal. For instance, one may consider the model with a 
direct coupling of the inflaton to the Einstein metric\footnote{We thank Gilles Esposito-Far\`ese for pointing out this possibility.}. 
Contrary to the situation with the universal coupling, now the scalar field $\varphi$ does not receive an effective tachyonic potential, and thus does not undergo the instability during inflation. Hence, one may naively expect that the model is viable even 
in the absence of the stabilizing potential $V(\varphi,\chi)$ introduced in Eq.~(\ref{inter}). In this case, however, the scalar $\varphi$ enjoys the shift symmetry, and hence can take on any value. Modulo fine-tuning, this value is 
not small, leading to a large value of $\varphi_0$ now and consequently to the conflict with Solar system tests. Moreover, even if the background value of $\varphi$ is tuned to zero, 
perturbations $\varphi$ are still too large and give rise to $\varphi_0 \gg 1$ (see the discussion in the first paragraph of Sec.~4.2). Once again, the problem is avoided upon turning on the potential $V(\varphi, \chi)$ as in Eq.~(\ref{inter}).

Yet another possibility is to couple the inflaton to the Einstein metric with a conformal factor as in Eq.~(\ref{coupling}), but with positive $\beta_{\text{inf}}>0$ (while at the same time keeping $\beta<0$ for the normal matter to ensure scalarization). 
In this case, according to Eq.~(\ref{tachyonic}), the field $\varphi$ acquires a positive mass even if $V(\varphi, \chi)=0$. 
Provided that $\beta_{\text{inf}}\gg 1$, the scalar $\varphi$ is superheavy. Thus it relaxes to zero exactly in the same way as in the model with the stabilizing potential $V(\varphi, \chi)$.
In fact, one can view this scenario as a variation of the model discussed in the main body of the paper, modulo the replacement of the coupling 
$\sim \varphi^2\chi^2$ by the coupling of the field $\varphi$ to the trace of the inflaton energy-momentum tensor.

It is worth to contrast the results of this paper on the modification of DEF model with those of Ref.~\cite{Anson:2019uto}, 
which raises a doubt in the validity of scalarization scenarios involving the Gauss-Bonnet curvature invariant.
The mechanism which relaxes the field $\varphi$ to zero presented here is not applicable to the Gauss-Bonnet case for the reason that 
the stabilization of the tachyonic mass during inflation would require the coupling $g^2$ in Eq.~(\ref{inter}) to be of order $10^{53}$~\cite{Anson:2019uto}. Such values of the dimensionless coupling constant would put the theory in the strong coupling regime. 

Note that the results of the present work are largely insensitive to the 
structure of the conformal factor $A(\varphi)$. While we have focused on the simple quadratic 
function $ \ln A(\varphi) \propto \varphi^2$, 
involving higher powers of $\varphi$ would leave our analysis and conclusions intact. 
Moreover, the proposed solution of taming the cosmological 
instability can be applicable to other models of scalarization akin to the DEF model. 
Indeed, starting from the action~(\ref{DEFEinstein}), one can make the disformal transformation of the metric as
$g_{\mu\nu}\to C(X)g_{\mu\nu}+D(X)\partial_\mu\varphi \partial_\nu\varphi$, where $C(X)$ and $D(X)$ are functions 
of the kinetic term $X=\left(\partial\varphi\right)^2$. 
The transformation results in a new scalar-tensor action~\cite{Zumalacarregui:2013pma}, belonging to the class of Degenerate Higher-Order Scalar-Tensor theories~\cite{Langlois:2015cwa}.
In the context of scalarization such extensions have been discussed in Refs.~\cite{Minamitsuji:2016hkk,Andreou:2019ikc}.
We believe that our solution for the cosmological instability presented in this paper may also work for such an extension.
However, the detailed analysis of this issue is beyond the scope of our paper.

\section*{Acknowledgments} 
E.B. is grateful to Gilles Esposito-Far\`ese for interesting discussions and critical reading of the manuscript. 
T. A. and E. B. acknowledge support from the research program “Projet 80|PRIME CNRS” and CNRS/RFBR Cooperation program 2018-2020 n. 1985.
S.R. is supported by the European Regional Development Fund (ESIF/ERDF) 
and the Czech Ministry of Education, Youth and Sports (M\v SMT) through the Project CoGraDS-CZ.02.1.01/0.0/0.0/15\_003/0000437. 

\section*{Appendix: evolution of the field $\varphi$ during inflation in the original DEF model}

In this Appendix, we discuss the inflationary evolution of the field $\varphi$ in the original DEF scenario. This evolution is subject to a tachyonic instability. As a result, the field $\varphi$ acquires large values 
inconsistent not only with the Solar system constraints, but also with the existence of the inflationary stage. This conclusion 
holds even if classically the field $\varphi$ is set exactly at $\varphi=0$ initially. Inevitable vacuum fluctuations of the 
field $\varphi$ are quickly enhanced during inflation leading to the large overall value of $\varphi$. In the following, we quantify the effect of vacuum fluctuations assuming 
the exact de Sitter approximation characterized by the Hubble expansion rate $H$. Switching to the canonically normalized field $\hat{\varphi}$ defined by Eq.~\eqref{canonical}, we write for perturbations $\delta \hat{\varphi}$ obeying the
Bunch--Davies vacuum initial conditions: 
\begin{equation}
\nonumber 
\delta \hat{\varphi}=\int \frac{\text{d}^3{\bf k}}{(2\pi)^{3/2}} \frac{\sqrt{\pi}}{2} H |\eta|^{3/2} \left[H^{(2)}_{\nu} (k|\eta |) e^{-i{\bf kx}}A^{\dagger}_{{\bf k}}+H^{(1)}_{\nu} (k|\eta|) e^{i{\bf kx}} A_{{\bf k}} \right] \; ,
\end{equation}
where $H^{(1,2)}_{\nu} (k|\eta|)$ are the Hankel functions of order $\nu=\sqrt{\frac{9}{4}+6 |\beta|}$, and $(A^{\dagger}_{{\bf k}}, A_{{\bf k}})$ is the pair 
of creation-annihilation operators. The expectation value of $\delta \hat{\varphi}$ is drawn from 
\begin{equation}
\nonumber 
\langle (\delta \hat{\varphi} )^2 \rangle =\int \frac{\text{d}k k^2}{8\pi} H^2 |\eta|^3 \left| H^{(1)}_{\nu} (k|\eta |)  \right|^2 \; .
\end{equation}
We are interested in superhorizon modes, i.e., $k |\eta| \rightarrow 0$, which add up to the classical background of the field $\hat{\varphi}$. In this limit, one has for the Hankel functions 
\begin{equation}
\nonumber 
H^{(1,2)}_{\nu} (k|\eta |) =\mp \frac{i \Gamma (\nu)}{\pi} \cdot \left(\frac{2}{k|\eta |} \right)^{\nu}  \; .
\end{equation}
The result reads 
\begin{equation}
\nonumber 
\langle (\delta \hat{\varphi} )^2 \rangle_{\{k \}}=\frac{2^{2\nu} \Gamma^2 (\nu) }{8(2\nu-3)\pi^3 } \cdot H^2 \cdot \left[(k_{\text{min}} |\eta|)^{3-2\nu}-(k_{\text{max}} |\eta|)^{3-2\nu} \right] \; .
\end{equation} 
Here $\{k \}$ denotes the range of momenta $(k_{\text{min}}, k_{\text{max}})$. Given that $\nu \simeq 5$ and assuming $k_{\text{max}} \gg k_{\text{min}}$, the second term in the square brackets is irrelevant. Conservatively, 
one can take $k_{\text{min}} \simeq H_0$ (we set the scale factor $a=1$ today) corresponding to the longest mode interesting in cosmology. The final expression in terms of the original field $\varphi$ is then 
given by 
\begin{equation}
\nonumber 
\langle (\delta \varphi)^2 \rangle_{\{k \}}=\frac{2^{2\nu} \Gamma^2 (\nu)}{2(2\nu-3)\pi^2 } \cdot \frac{H^2}{M^2_{Pl}} \cdot \left| \frac{\eta_{*}}{\eta}\right|^{2\nu-3} \; ,
\end{equation}
where $\eta_*$ denotes the time when the cosmological mode with wavenumber $k_{\text{min}}$ exits horizon. It is evident that $\sqrt{\langle (\delta \varphi)^2 \rangle_{\{k \}}}$ is very large for $|\eta_*| \gg |\eta|$. Given the minimal duration of inflation, which should last for at least $50-70$ e-foldings, 
we end up with an unacceptably large $\varphi$. This huge $\varphi$ clearly violates existing Solar system constraints, and also threatens the existence of the inflationary stage. 

One solution to this problem is discussed in the main body of the paper. That is, to equip the field $\varphi$ with the coupling to the inflaton. Another approach is to take into account higher order terms in the function $\ln A(\varphi)$. Namely, if a quartic term is present in the expansion of $\ln A(\varphi)$, as is written in Eq.~\eqref{quartic}, the field $\varphi$ rolls towards its minimum set at $\varphi =\sqrt{-\beta/\lambda}$ and resides there up to the present day. 
However, in this latter approach with $\lambda >0$, a non-zero cosmological value of $\varphi$ is inconsistent with the scalarization of neutron stars~\cite{Anderson:2016aoi}.

\end{document}